\documentclass[useAMS,letter]{emulateapj}

\usepackage{amssymb}
\usepackage{natbib}
\usepackage{graphicx}

\def\pc{\thinspace\mathrm{pc}}     
\def\K{\thinspace\mathrm{K}}     
\def\dens{\thinspace m_\mathrm{p}\thinspace\mathrm{cm}^{-3}} 
\def\ndens{\thinspace\mathrm{cm}^{-3}} 
\def\Myr{\thinspace\mathrm{Myr}}
\def\kyr{\thinspace\mathrm{kyr}}
\def\Msun{\thinspace M_\odot}
\def\Lyunit{\thinspace\gamma\thinspace\mathrm{cm}^{-2}\thinspace\mathrm{s}^{-1}}
\def\kms{\thinspace\mathrm{km}\thinspace\mathrm{s}^{-1}}
\def\sdens{\thinspace\mathrm{cm}^{-2}}
\def\cm{\thinspace\mathrm{cm}}

\def\Simfid{{\it Mach 5 (G09b)}}
\def\SimMvl{{\it Mach 1.5}}
\def\SimMvh{{\it Mach 12.5}}
\def\Simlr{{\it low resolution}}
\def\Simopen{{\it open boundaries}}
\def\Simwif{{\it Mach 5 warm}}
\def\Simwhf{{\it M5 warm high flux}}
\def\Simlf{{\it low flux}}
\def\Simhf{{\it high flux}}
\def\Simll{{\it low density}}
\def\Simk4{{\it smaller $k_{\mathrm{max}}$}}
\def\SimMl{{\it Mach 4}}
\def\SimMh{{\it Mach 7}}
\def\Simsb{{\it small box}}
\def\Simbb{{\it big box}}

\shorttitle{Detailed Numerical Simulations on the Formation of Pillars}
\shortauthors{Gritschneder et al.}

\begin{document}

\title{Detailed Numerical Simulations on the Formation of Pillars around
HII-regions}

\author{Matthias Gritschneder$^{1,2}$\footnote{gritschneder@pku.edu.cn}, Andreas
  Burkert$^{2,3}$\footnote{Max Planck Fellow}, Thorsten
  Naab$^{2,4}$, Stefanie Walch$^{5}$}
\affil{$^1$ Kavli Institute for Astronomy and Astrophysics, Peking University,
Yi He Yuan Lu 5, Hai Dian, 100871 Beijing, China}
\affil{$^2$ Universit\"ats-Sternwarte M\"unchen, Scheinerstr.\ 1,
  81679 M\"unchen, Germany }
\affil{$^3$ Max Planck Institut f\"ur Extraterrestrische Physik,
  Giessenbachstr., 85748 Garching bei M\"unchen, Germany}
\affil{$^4$ Max Planck Institut f\"ur Astrophysik,
  Karl Schwarzschild Str. 1, 85740 Garching bei M\"unchen, Germany}
\affil{$^5$ School of Physics \& Astronomy, Cardiff University, 5 The
  Parade, Cardiff CF24 3AA, United Kingdom}

\begin{abstract}
We study the structural evolution of turbulent molecular clouds under
the influence of ionizing radiation emitted from a nearby massive star
by performing a high resolution parameter study with the iVINE code.
The temperature is taken to be $10\K$ or $100\K$, the mean number density is either
$100\ndens$ or $300\ndens$. Furthermore, the turbulence is varied between
Mach $1.5$ and Mach $12.5$, the main driving scale of the turbulence
is varied between $1\pc$ and $8\pc$. We vary the ionizing flux by an
order of magnitude, corresponding to allowing between $0.5\%$ and
$5\%$ of the mass in the domain to be ionized immediately.
In our simulations the ionizing radiation enhances the initial turbulent
density distribution and thus leads to the formation of
pillar-like structures observed adjacent to HII regions in a natural
way. Gravitational collapse occurs regularly at the tips of the structures. 
We find a clear correlation between the initial state of the turbulent
cold cloud and the final morphology and physical properties of the
structures formed. 
The most favorable regime for the formation of
pillars is Mach $4-10$. Structures and therefore stars only form
if the initial density contrast between the high density unionized gas
and the gas that is going to be
ionized is lower than the temperature contrast
between the hot and the cold gas.
The density of the resulting pillars is determined by a pressure
 equilibrium between the hot and the cold gas. 
A thorough analysis of the simulations shows that the complex
kinematical and geometrical structure of the formed elongated
filaments reflects that of observed pillars to an impressive level of
detail. 
In addition, we find that the observed line-of sight velocities allow
for a distinct determination of different formation
mechanisms. Comparing the current simulations to previous results and
recent observations we conclude that e.g. the
 pillars of creation in M16 formed by the mechanism proposed here and
 not by the radiation driven implosion of  pre-existing clumps.
\end{abstract}


\section{Introduction}
Stars are known to form in turbulent, cold molecular clouds. When massive stars
ignite, their UV-radiation ionizes and heats the surrounding gas, leading to an expanding
HII bubble. As soon as the HII region breaks through the surface of
the molecular cloud a low-density, optically thin hole is formed,
which reveals the otherwise obscured interior. 
At the interface between the HII region and the molecular gas
peculiar structures, often called pillars, are found. 
The most famous examples are the 'pillars of creation' in 
the Eagle Nebula (M16, \citealt{1996AJ....111.2349H}). There is also
wide-spread evidence for star formation at the
tips of the pillars \citep[e.g.][]{2002ApJ...565L..25S,
 2002ApJ...570..749T,2007PASJ...59..507S,2008hsf2.book..599O}.
Since the launch of the Spitzer Space Telescope a wealth of highly resolved
observations of the peculiar, pillar or trunk like structures observed
around the hot, ionized HII-regions around massive stars and the star
formation in this trunks has become available, e.g. in the Orion clouds
\citep{2002A&A...393..251S, 2007ApJ...657..884L, 2009AJ....137.3685B}, the 
Carina nebula \citep{2000ApJ...532L.145S}, the Elephant Trunk Nebula
\citep{2004ApJS..154..385R}, the Trifid Nebula
\citep{2002ApJ...581..335L}, the Rosette
Nebula \citep{Schneider:2010fk}, M16 \citep{2004A&A...414..969A}, M17   
\citep{2002ApJ...577..245J}, 30 Dor \citep{2002AJ....124.1601W} and
the SMC \citep{2007ApJ...665..306G}.
In addition, several recent observations of bright rimmed clouds 
(BRCs) \citep{2009arXiv0902.4751U,
  2009arXiv0903.2122C,Morgan:2009ys,Morgan:2010lr} have been 
carried out.
An interesting aspect is the surprisingly spherical shape of many
observed nebulae, especially in RCW 120, 'the perfect bubble'
\citep{2009A&A...496..177D,Zavagno:2010vn}. Other regions, like e.g. RCW 79
\citep{2006A&A...446..171Z}, RCW82 \citep{2009A&A...494..987P}, RCW
108 \citep{2007A&A...473..149C} and Sh 104
\citep{Deharveng:2003yq,Rodon:2010kx} share this morphology. 

In general, the pillars point like fingers towards the ionizing source
and show a common head-to-tail 
structure. Most of the mass is concentrated in the head which has a
bright rim facing the young stars \citep[e.g.][]{2006A&A...454..201G}. 
Thin, elongated pillars connect the head with 
the main body of the molecular cloud. They have typical widths 
of $0.1-0.7\pc$ and are $1-4\pc$
long \citep{2006A&A...454..201G,2006A&A...454L..87S}. 
The observations show that the pillars are not smooth, but show small
scale structure, filaments and clumps \citep{1998ApJ...493L.113P}. Some
filaments run diagonal across the pillars, suggesting a complex twist into a helical
structure \citep{2003A&A...403..399C}. 
This is also supported by spectroscopic measurements of the
line-of-sight (LOS) gas velocity: the pillars show a bulk motion
away from the ionizing stellar sources
with a superimposed complex shear flow that could be interpreted as
corkscrew rotation \citep{2006A&A...454..201G}. 
Occasionally, close to the
tip of the head small spherical gas clumps are observed to break off
and float into the hot HII region.
These so called evaporating gaseous globules (EGGs) have been found with
HST e.g. in the Eagle Nebula \citep{2002A&A...389..513M}. If stars with
surrounding gas discs happen to form in these clumps they transform
into evaporating  proto-planetary discs, so called
proplyds. More recent observations \citep{Gahm:2007lr} have revealed a wealth of even
smaller sized globules or globulettes, which are in general not bright
rimmed and show no detectable sign of star formation.
Direct signatures of star formation are found in the head of 
the pillars, e.g. through jets from obscured proto-stars piercing through the
surface into the HII region (e.g. in Eta Carina,
\citealt{2000ApJ...532L.145S}). Whether these jets are  preferentially aligned
perpendicular to the trunk is a matter of current debate
\citep{Raga:2010qy,Smith:2010fk}.

On the theoretical side, early models of pillar formation suggested that 
they form by Rayleigh-Taylor instabilities when the
expanding hot, low-density HII region radially accelerates
the cold, dense gas \citep{1954ApJ...120...18F}.
This has been ruled out by the observations of the complex flows
inside the pillars \citep{1998ApJ...493L.113P}.

Another possibility is the collect and collapse model. Here, the
radiation sweeps up a large shell, which then fragments to form stars
and
pillars \citep{1977ApJ...214..725E,1980SSRv...27..275K,1982ApJ...260..183S}. However,
the timescales ($>5\Myr$) and masses ($>1000\Msun$)
involved \citep{1995ApJ...451..675E,2001A&A...374..746W} are much larger
than in M16. Therefore, this is a more likely scenario for supernova-driven shells.

A third scenario is the radiation-driven
implosion (RDI, e.g. \citealt{1989ApJ...346..735B}) of pre-existing dense cores. This
has been studied in great detail with numerical simulations
\citep[e.g.][]{1994A&A...289..559L,2001MNRAS.327..788W}. More recently \cite{2003MNRAS.338..545K} presented three-dimensional 
RDI simulations with a smoothed-particles-hydrodynamics (SPH) code and were able to show that an
otherwise gravitationally marginally stable sphere can be driven into
collapse by ionizing radiation. In \citet[][ hereafter G09a]{2009MNRAS.393...21G} we showed
that marginally stable density enhancements 
get triggered into forming stars in cases with high as well as
low ionizing flux. \cite{2009ApJ...692..382M} further analyzed this RDI-scenario
with a SPH-based radiative transfer scheme. They show that
there is an evolutionary sequence, depending on the initial size of
the cloud, as suggested by
\cite{1994A&A...289..559L}. \citet{2009A&A...497..649B} studied the
implosion of a single clump with a new ray-tracing scheme, based on
the HEALPix algorithm. 
An new implementation in the adaptive-mesh-refinement (AMR) code FLASH using the hybrid
characteristics raytracing was achieved by
\citet{2009AAS...21344103P,Peters:2010uq}. Very recently \citet{Mackey:2010uq} were 
able to reproduce the density structure of the main pillar in M16 by
setting up several pre-existing dense cores in a triangular way. They
investigate the effects of different cooling recipes in a grid code but are missing
the effects of self-gravity.

Recently, the focus has moved towards the ionization of the turbulent
ISM. \cite{2006ApJ...647..397M} reproduced
the observed morphologies of HII regions by ionizing a turbulent
medium using a grid code without the inclusion of
gravity. \cite{2007MNRAS.375.1291D} 
used an SPH code to compare the gravitational collapse of a molecular
cloud with and without ionization. They found slightly enhanced star 
formation in the simulation with ionization. The inclusion of
ionization in a grid code in combination with a
magnetic field  was discussed by \cite{2007ApJ...671..518K}. A
homogenous magnetic field leads to a non-spheric HII-region, as the gas
is held back by the magnetic field lines and an oval shaped bubble
develops. 
However, all these simulations focussed on the evolution of the entire HII-region
and therefore lack the resolution for a detailed kinematical and
structural analysis of the pillars. For a comprehensive quantitative
study we investigated the radiative ionization of
a turbulent molecular cloud from a nearby star cluster with so far
unprecedented resolution by zooming into a
subregion of the cloud. 
First, we focussed on the evolution of the power spectrum in the 
interaction zone of a turbulent cloud affected by ionization \citep[][
hereafter G09b]{2009ApJ...694L..26G}. Recently, 
\citet{Lora:2009yq} further investigated the ionization of a
turbulent cloud by the combination of a two-temperature equation of
state with gravitational forces and transfer of ionizing
radiation. They produce pillar-like structures including cores. The angular
momentum of these cores is preferentially aligned perpendicular to the
direction of the ionizing field.

A different approach was presented by \citet{Nayakshin:2009rt}. They
focussed on the momentum transfer in regimes of high radiation
pressure by combining SPH with a Monte-Carlo approach.
Another main motivation of developing software able to treat ionizing radiation
have been investigations of the re-ionization of the early universe
\citep[see e.g.][ and references therein]{2006MNRAS.371.1057I,
  2008MNRAS.389..651P, 2008MNRAS.386.1931A,Iliev:2009kx}.

In this work we investigate for the first time the effect of different
initial conditions on the ionization of turbulent molecular clouds. We
vary the initial temperature, the mean number density, the level of
turbulence as well as the turbulent scale and the ionizing flux.
The structure of this paper is
as follows. In \S \ref{Basics} we briefly review the concept of
ionizing radiation, followed by a short summary of the
iVINE-code. After that we present the set of initial conditions for the
parameter study. In \S \ref{Results} the outcome of the different
simulations is discussed in detail. \S \ref{Globules} is dedicated to
the stability and shape of the structures in dependance on the initial conditions.
A close comparison to the observed masses, morphologies and line-of-sight
(LOS) velocities is done in \S \ref{Comp_Obs}. We draw the conclusions
in \S \ref{Conclusions}.

\section{Basic Approach and Initial Conditions}
\label{Basics}
\subsection{Ionizing Radiation}
As soon as an O star is born it ionizes its surroundings with its
UV-radiation. This leads to an ionized, hot HII-region
($T_\mathrm{ion}\approx10^4 \K$). In the beginning the 'R-type' 
ionization front travels with a speed $v_\mathrm{R}$ which is larger
than the sound speed of the hot gas $a_\mathrm{ion}$. This phase ends
as soon as ionization is balanced by recombination in the
HII-region. The ionized volume $V_\mathrm{S}$, the so called
Str{\"o}mgren sphere \citep{1939ApJ....89..526S}, is then given by 
\begin{equation}
\label{Vol_Stroem}
V_\mathrm{S}=\frac{J_\mathrm{Ly}}{\alpha_\mathrm{B}(\rho_0/m_p)^2}.
\end{equation}
Here, $J_\mathrm{Ly}$ is the flux of ionizing photons of the source,
which is assumed to be constant and monochromatic, $\alpha_\mathrm{B}$
is the recombination coefficient, $\rho_0$ is the density of the
pre-existing, homogeneous gas and 
$m_\mathrm{p}$ the proton mass. At a larger distance from the star the
radiation can be assumed to impinge onto a surface in a plane-parallel way and
thus Eq. \ref{Vol_Stroem} simplifies to 
\begin{equation}
\label{x_Stroem}
x_\mathrm{S}=\frac{F_\mathrm{Ly}}{\alpha_\mathrm{B}(\rho_0/m_\mathrm{p})^2},
\end{equation}
where $x_\mathrm{S}$ is the thickness of the ionized region and
$F_\mathrm{Ly}$ is the in-falling ionizing flux per unit time and unit
area. 

After a hydrodynamical timescale the hot gas reacts to its
increased temperature and pressure. The pressure of an ideal
one-atomic gas is given by
\begin{equation}
\label{P_iso}
P=\rho\frac{k_\mathrm{B}T}{\mu m_\mathrm{p}}=\rho c_\mathrm{s}^2,
\end{equation}
where $\rho$ is the density, $k_\mathrm{B}$ is the Boltzmann constant,
$\mu$ is the mean molecular weight and $c_\mathrm{s}$ is the isothermal
sound-speed. Now the evolution is characterized by 
an isothermal shock followed by a weaker, 'D-type' ionization
front. The front velocity is now $v_\mathrm{D}<a_\mathrm{ion}$. For a
full analysis see e.g. \citet{1991pagd.book.....S}. As the
hot gas expands, its density is reduced. At the same time, the cold
surrounding gas is compressed. Under the assumption that the
homogeneously ionized region consumes all UV-photons of the source it
follows from Eq. \ref{x_Stroem} that the density of the hot gas for a
constant flux and a constant temperature  $T_\mathrm{hot}$  at any given
time is
\begin{equation}
\label{rho_Stroem}
\rho_\mathrm{hot}(t) = \sqrt{\frac{m_P^2 F_\mathrm{Ly}}{\alpha_\mathrm{B}x(t)}}.
\end{equation}
To calculate the front position $x(t)$ we follow the approach of
\cite{2003adu..book.....D}. Under the assumption of a thin shock which
is traveling at a speed $v_\mathrm{s}$ the
ram pressure in the hot, ionized gas has to be equal to the ram
pressure in the cold gas
\begin{equation}
P_\mathrm{ion}=P_\mathrm{cold},
\end{equation}
where 
\begin{equation}
\label{P_cold}
P_\mathrm{cold}=\rho_\mathrm{0}\,v_\mathrm{s}^2=\rho_\mathrm{0}
\left(\frac{dx}{dt}\right)^2 .
\end{equation}
The pressure on the ionized side of the shock is mainly given by the thermal
pressure of the hot gas
\begin{equation}
\label{P_ion}
P_\mathrm{ion}=fP_\mathrm{hot}=f\rho_\mathrm{hot}c_\mathrm{s,hot}^2,
\end{equation}
where we have already introduced a constant fitting factor $f$ to
account for the approximations made (e.g.  the one leading to
Eq. \ref{rho_Stroem}). Combining Eq. \ref{P_cold} and \ref{P_ion} and using
Eqs. \ref{x_Stroem} and \ref{rho_Stroem} yields
\begin{equation}
x^\frac{1}{4}\frac{dx}{dt}=fc_\mathrm{s,hot}\,x_\mathrm{s}^\frac{1}{4}.
\end{equation}
With the initial condition $x(t_0)=x_\mathrm{s}$ we can solve by integration and
obtain
\begin{equation}
\label{x_front}
x(t)=x_\mathrm{s}\left(1+f\frac{5}{4}\frac{c_\mathrm{s,hot}}{x_\mathrm{s}}(t-t_0)\right)^\frac{4}{5}.
\end{equation}
Using Eq. \ref{rho_Stroem} it follows that
\begin{equation}
\label{rho_front}
\rho_\mathrm{hot}(t)= \rho_0\left(1+f\frac{5}{4}\frac{c_\mathrm{s,hot}}{x_\mathrm{s}}(t-t_0)\right)^{-\frac{4}{10}}
\end{equation}
for a plane-parallel infall of a constant flux onto a homogeneous medium.

\subsection{Numerical Method and First Tests}
\label{method}
In order to investigate the effect of different initial conditions and 
levels of UV-radiation on the formation of pillars we conduct a parameter
study. All simulations were performed with iVINE (G09a), an
implementation of ionizing radiation in the tree-SPH code VINE
\citep{2009ApJS..184..298W,2009ApJS..184..326N}. 
As the mean free path of the atoms and electrons in the cold and the
hot phase is of order $\lambda\approx10^{12}\cm$ and
$\lambda\approx10^{14}\cm$, respectively (see
e.g. \citet{1991pagd.book.....S}), which is much smaller than the
lenght-scales involved in our simulations, the gas can be treated by the means
of fluid dynamics. The evolution of a turbulent molecular cloud under
the influence of ionization spans several orders of magnitude in
density. We therefore chose to solve the hydrodynamic
equations with the method SPH, a Lagrangian approach with adaptive resolution.
To prevent artificial fragmentation \citep{1997MNRAS.288.1060B} the
Jeans length has always to be resolved with at least 50
particles. This leads to a resolution limit which we ensure to be
small enough by using a sufficient amount of total particles (see \S
\ref{RES_SF}).

In iVINE, the ionizing
radiation is assumed to impinge plane-parallel onto the simulated
volume from the negative x-direction. From the surface of infall the
radiation is propagated along the x-direction by a ray-shooting
algorithm. Along these rays the ionization degree $\eta_\mathrm{i}$ is
calculated for each particle i. According to the ionization degree
the pressure $P_\mathrm{i}$ of the particle is calculated by a linear
interpolation between the temperature $T_\mathrm{hot}$ of the hot, ionized
and the temperature $T_\mathrm{cold}$ of the cold, un-ionized
gas. Here, we assume both gas components to be isothermal, since for
the density range considered in our simulations heating and cooling should
balance each other to approximate isothermality \citep[see
e.g.][]{1998ApJ...504..835S}. Following Eq. \ref{P_iso} the new pressure in our simulation is
given as 
\begin{equation}
P_\mathrm{i} = \left(\frac{T_\mathrm{ion} \eta_\mathrm{i}}{\mu_\mathrm{ion}} +
  \frac{T_\mathrm{nion} (1-\eta_\mathrm{i})}{\mu_\mathrm{nion}}\right)
\frac{k_\mathrm{B} \rho_\mathrm{i}}{m_\mathrm{p}},
\end{equation}
where $\rho_\mathrm{i}$ is the SPH-density of the particle $i$ and
$\mu_\mathrm{ion}=0.5$ and $\mu_\mathrm{nion}=1.0$ are the mean
molecular weights of the ionized and the un-ionized gas in the case of
pure hydrogen, respectively. 

As a first test we verify Eq. \ref{x_front} and fit a value
for $f$ by ionizing a slab of atomic hydrogen with a constant
homogeneous density of $\rho_\mathrm{cold}=300\dens$ and temperature of
$T_\mathrm{cold}=10\K$. We perform three different runs, corresponding to a low
flux ($F_\mathrm{Ly}=1.66\times 10^9\Lyunit$), an intermediate flux
($F_\mathrm{Ly}=5\times 10^9\Lyunit$) and a high flux
($F_\mathrm{Ly}=1.5\times 10^{10}\Lyunit$). This corresponds to the 
ionization penetrating immediately into the first  $0.55\%$, $1.67\%$ and
$5\%$ of the region, respectively (see Eq. \ref{x_Stroem}). At the
same time this is equal to placing the simulation volume further away
or closer to the source,  e.g. the O-star. The simulations are
conducted with the same accuracy and setup as in the parameter study
given below (see \S \ref{ICs}).
\begin{figure}
  \centering{
   \plotone{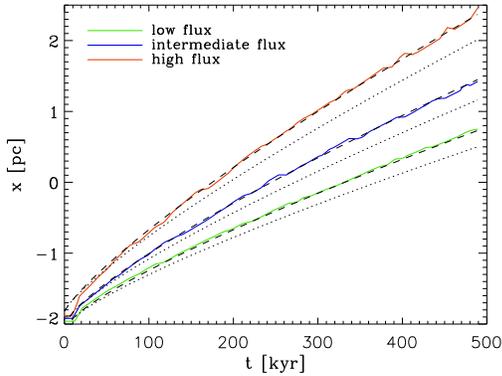}
  \caption{Front position versus time for the three test simulations
    with a different flux impinging on a homogeneous medium. Green,
    blue and red line: simulations with a low, intermediate and high
    flux, respectively. Black lines: solution according to
    Eq. \ref{x_front}, dotted $f=1$, dashed $f=\sqrt{\frac{5}{4}}$.\label{frontspeed}}}
\end{figure}
Fig. \ref{frontspeed} shows the resulting evolution of the front. As
one can clearly see the approximations leading to Eq. \ref{x_Stroem}
($f=1$, the dotted lines in Fig. \ref{frontspeed}) do not produce
satisfactory results. Instead, assuming
\begin{equation}
P_\mathrm{ion}=\sqrt{\frac{5}{4}}P_\mathrm{hot}
\end{equation}
(i.e. $f=\sqrt{\frac{5}{4}}$, the dashed lines in Fig.\ref{frontspeed})
perfectly matches the simulations during the entire simulated time of
$t_\mathrm{sim}=500\kyr$. Thus, we keep this assumption for this work.

\subsection{Initial Conditions}
\label{ICs}
To produce different turbulent initial conditions we use the same
approach as in G09b. We set up $2\times 10^6$ particles to resemble a
homogeneous medium with $\rho_\mathrm{cold}=300\dens$ in a volume of
$(4\pc)^3$. Then, we add a supersonic 
velocity field (Mach 10 in most cases) with a steep power-law $E(k)\propto
k^{-2}$ on the modes $k=1..4$. Before switching on the ionizing source
each setup decays 
freely under the influence of isothermal hydrodynamics and periodic
boundary conditions until the desired initial Mach number is reached
(after $t\approx 0.8-1.0 \Myr$, depending on the specific
simulation). This self-consistent evolution of the turbulence leads to
a combination of solenoidal and compressible modes, which seems to be
the case in turbulent molecular clouds\footnote{For a more detailed
  investigation of the role of compressional and solenoidal modes see
  \citet{Federrath:2010fj}}.
The individual particle time-steps in iVINE are determined as in G09b by using an
accuracy parameter of $\tau_\mathrm{acc}=1.0$ and a
Courant-Friedrichs-Lewy (CFL) tolerance parameter of $\tau_\mathrm{CFL}=0.3$. 
An additional time-step criterion based on the maximum allowed change of
the smoothing length with an accuracy parameter of
$\tau_\mathrm{h}=0.15$ is also employed.

\begin{table*}
\begin{center}
\begin{tabular}{lcccccccc}
\hline
Simulation & $M$ &  $\bar{\rho}$ & 
$l_\mathrm{box}$ &
$F_\mathrm{Ly}$ & Mach & $k$ &
$ T_\mathrm{nion}$ & $log_{10}(N_\mathrm{part})$\\
 & $[\Msun]$ &  $[\dens]$ & $[\pc]$ & $[\Lyunit]$ & & &
$[\K]$ & \\
\hline
\Simlr & 474 & 300 & 4 & $5\times 10^9$ & 5 & 1-4 & 10 & 5.4\\
\Simopen & 474 & 300 & 4 & $5\times 10^9$ & 5 & 1-4 & 10 & 6.3 \\
\SimMvl & 474 & 300 & 4 & $5\times 10^9$ & 1.51 & 1-4 & 10 & 6.3 \\
\SimMl & 474 & 300 & 4 & $5\times 10^9$ & 4 & 1-4 & 10 & 6.3 \\
\Simfid & 474 & 300 & 4 & $5\times 10^9$ & 5 & 1-4 & 10 & 6.3 \\
\SimMh & 474 & 300 & 4 & $5\times 10^9$ & 7 & 1-4 & 10 & 6.3 \\
\SimMvh \footnote{In the case \SimMvh\, we start with an initial
  velocity field of Mach 20 which is then allowed to decay freely as
  prescribed in \S\ref{ICs}} 
& 474 & 300 & 4 & $5\times 10^9$ & 12.5 & 1-4 & 10 & 6.3 \\
\Simwif & 474 & 300 & 4 & $5\times 10^9$ & 5 & 1-4 & 100 & 6.3 \\
\Simwhf & 474 & 300 & 4 & $1.5\times 10^{10}$ & 5 & 1-4 & 100 & 6.3 \\
\Simlf & 474 & 300 & 4 & $1.7\times 10^9$ & 5 & 1-4 & 10 & 6.3 \\
\Simhf & 474 & 300 & 4 & $1.5\times 10^{10}$ & 5 & 1-4 & 10 & 6.3 \\
\Simll & 158 & 100 & 4 & $ 1.7\times 10^9$ & 5 & 1-4 & 10 & 6.3 \\
\Simk4 & 474 & 300 & 4 & $5\times 10^9$ & 5 & 4-8 & 10 & 6.3 \\
\Simsb\footnote{To allow for comparable results the box is 2pc in the y- and
  z-direction but 4pc in the x-direction. Therefore, the particle
  number is increased} 
 & 119 & 300 & 2 & $5\times 10^9$ & 5 & 1-4 & 10 & 6.6 \\
\Simbb & 3795 & 300 & 8 & $5\times 10^9$ & 5 & 1-4 & 10 & 6.3 \\
\hline
\end{tabular}
\caption{Listing of the different initial conditions. Given are
  initial mass, average density and size of the simulation. In
  addition, the impinging flux, turbulent Mach number, the largest
  driving mode of the turbulence and the temperature are
  listed. \Simfid\, is the standard case as presented in G09b.\label{IC_tab}}
\end{center}
\end{table*}

After the desired Mach number is reached, the ionization is turned
on. Now, the boundaries are still periodic in the y- and z-direction. In the negative
x-direction the boundary is reflecting to represent conservation of
flux towards the star, whereas in the positive x-direction the gas is
allowed to stream away
freely. To test this approach we perform one simulation with open
boundaries in both x-directions. Gravitational forces are calculated 
without boundaries. This is valid as the free-fall time
of the whole simulated area is $ t_\mathrm{ff}\approx 3 \Myr$, which is
much longer than the simulation time of $t_\mathrm{final} = 0.5
\Myr$. For the tree-based
 calculation of gravitational forces we use a multi-pole acceptance
 criterion (MAC, \citealt{2001NewA....6...79S}) with a tree accuracy
 parameter of $\theta=5\times10^{-4}$. The correct treatment of the
 ionization and the resulting acceleration of the particles is
 guaranteed by the modified CFL-condition discussed in G09a. The
 recombination of the hot gas is modeled assuming
 $\alpha_\mathrm{B}=2.59\times 10^{-13}$ and the cross-section for the
 ionizing photons is set to
 $\sigma=3.52\times10^{-18}\thinspace\mathrm{cm}^2$. The initial
 properties of the different simulations are listed in Table \ref{IC_tab}. The simulations
 were performed on a SGI Altix 3700 Bx2 supercomputer, the calculation
 of each setup took approximately 100 wall clock hours on 16 CPUs.

\section{Results of the Parameter Study}
\label{Results}
\subsection{General Properties}
\label{RES_general}
\begin{figure*}
  \centering{
   \includegraphics[width=19cm]{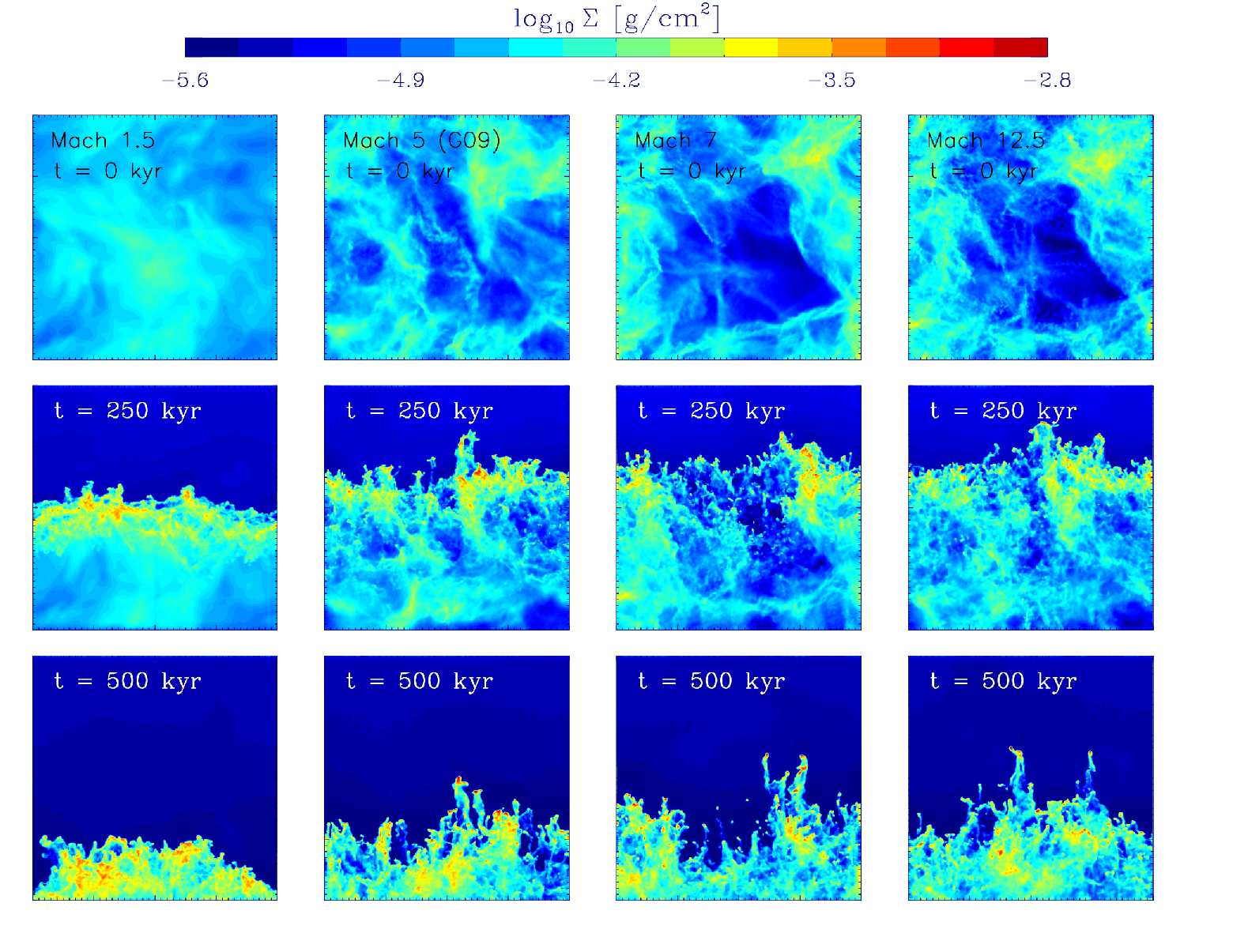}
  \caption{Time evolution of four different initial conditions. We
    show the density projected along the z-axis of four different simulations
    at subsequent stages. From left to right column by increasing Mach
    number, as indicated. Colour-coded is the surface density, each
    figure is $4\pc \times 4\pc$. The significant structures only form
    above a certain level of turbulence ($M\geq 2$) and get less
    stable with increasing Mach number. \label{FIG_evol}}}
\end{figure*}

\begin{figure*}
  \centering{
   \includegraphics[width=19cm]{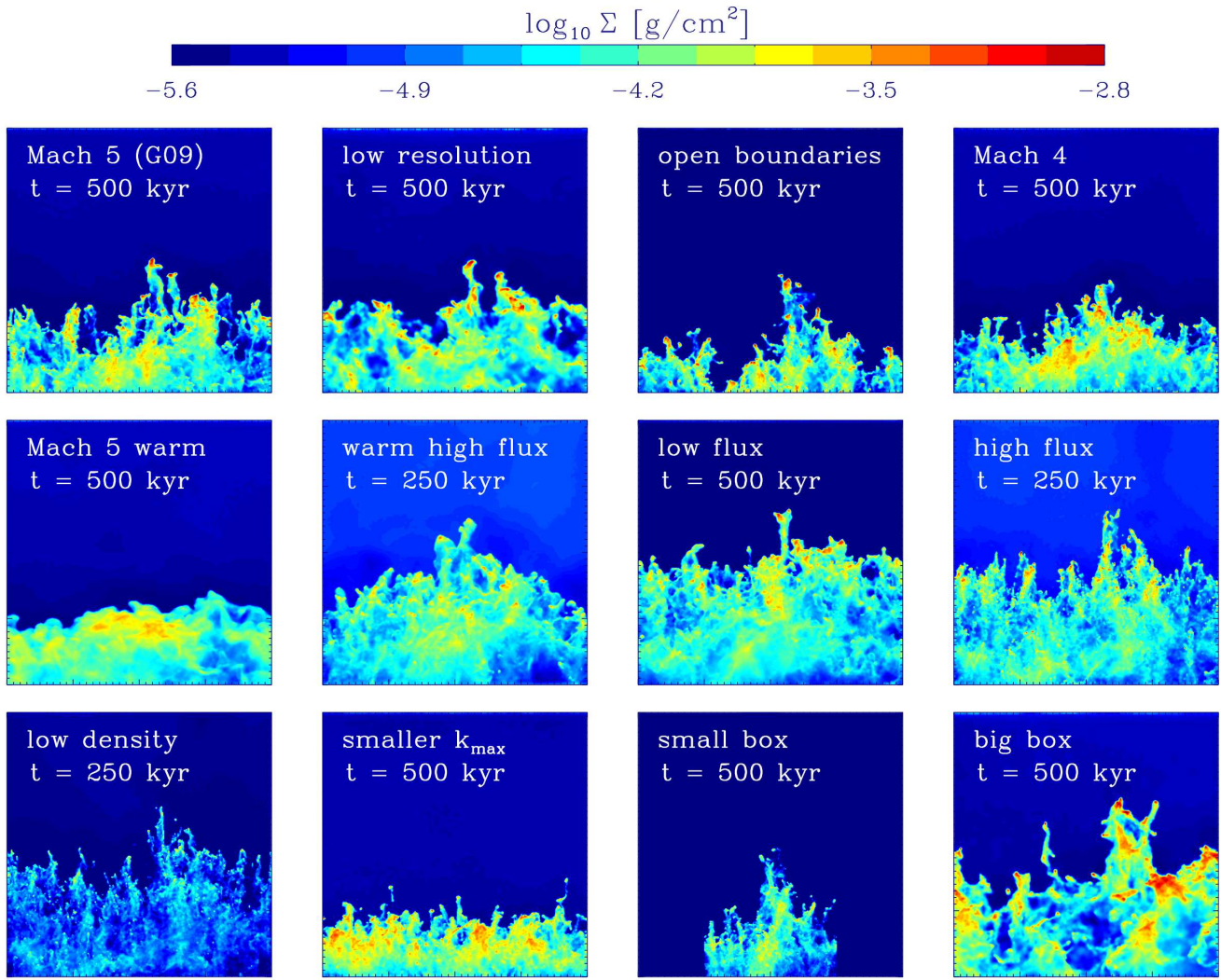}
  \caption{The final stage of the different simulations. Color-coded
    is the surface density of the simulations given in Table
    \ref{TAB_compare} as indicated in the top left corner, each
    figure is $4\pc \times 4\pc$. Denote that in
    panel 6, 8 and 9 an earlier stage is depicted since the evolution
    of the simulation is more rapid.\label{FIG_compare}. Since all
    panels have the same physical size panel 12 shows only the top
    left quarter of that simulation which includes the most
    significant structure.}}
\end{figure*}

The time evolution of the density for turbulent regions with Mach
 numbers of $1.5$, $5$, $7$ and $12.5$, respectively, is shown in Fig. \ref{FIG_evol}. 
In all of the simulations forming structures the same effect as
described in G09b takes place. The ionizing radiation penetrates
deeper into the turbulent cloud along low-density channels. As the
ionized gas reacts to its increase in pressure it starts to compress
the adjacent, un-ionized, higher density regions, thereby widening
the channels of low density and thus allowing the ionization to
penetrate even further. We call this process 'radiative round-up'. At
$t=250\kyr$ the pre-existing high-density structures have been
enhanced by the outside compression 
and pillars start to become visible. After $t=500\kyr$ the pillars have
achieved characteristic shapes 
which match the observations remarkably well. Fig. \ref{FIG_evol} (row
3) and Fig. \ref{FIG_compare} show the density projected along the z-axis
at this stage for all simulations of the parameter study.

\begin{table*}
\begin{center}
\begin{tabular}{lccccccccc}
\hline
Simulation & $M$ &  $\bar{\rho}_\mathrm{p}$ & 
$\bar{\Sigma}$ &  $\mathrm{log}_{10}[N(\mathrm{H_2})$ &
$\sigma$ &  
$\bar{v}_\mathrm{x}$ & $d$ &
$\bar{\rho}_\mathrm{ion}$ & $\Delta P$ \\
 & $[\Msun]$ &  $[10^{4}\dens]$ & 
$[10^{-3}\mathrm{g cm}^{-2}]$ & $\sdens]$ &
<\, $ [\kms] $ \, &  
\, $ [\kms] $ \, & $[\pc]$ &
$[\dens]$ & \\
\hline
\Simlr & $14.26$ & $5.66$  & $1.69$ & $20.5$ &
$0.9\pm0.5$ & $4.8\pm0.5$ & 0.11 &  42.5 & 1.50 \\
\Simopen  & $11.0$ & $5.06$  & $1.44$ & $20.5$ &
$1.2\pm0.7$ & $5.4\pm1.1$ & 0.10 &  28.6 & 1.13\\
\SimMvl & $6.2$ & $3.46$  & $1.18$ & $20.4$ &
$1.6\pm1.7$ & $4.5\pm1.5$ & 0.10  & 41.8 & 2.42  \\ 
\SimMl & $12.6$ & $3.62$  & $1.32$  & $20.4$ &
$1.3\pm0.6$ &  $3.9\pm0.9$ & 0.13& 39.8 & 2.20\\
\Simfid\, $250\kyr$ & $17.0$ & $3.60$  & $1.43$ & $20.5$ &
$1.8\pm1.7$ &  $4.0\pm1.7$ & 0.16 &  60.6 & 3.46\\
\Simfid & $12.6$ & $4.56$  & $1.52$ & $20.5$ &
$1.1\pm0.7$ &  $4.8\pm0.9$ & 0.12 &  43.4 & 1.90\\
\SimMh & $14.0$ & $5.36$  & $1.71$ & $20.6$ &
$1.1\pm0.5$ & $4.0\pm0.9$ & 0.11 & 39.2  & 1.46 \\ 
\SimMvh & $8.1$ & $5.66$  & $2.56$ & $20.7$ &
$1.3\pm1.9$ & $4.0\pm1.7$ & 0.09 & 43.5  & 1.54 \\ 
\Simwif & $10.2$ & $0.37$  & $0.23$ & $19.7$ &
$2.0\pm0.8$ & $3.3\pm1.7$ & - & 43.6 & 2.36\\
\Simwhf\, $250\kyr$  & $7.1$ & $2.59$  & $1.0$ & $20.3$ &
$3.5\pm3.6$ & $7.0\pm3.7$ & 0.11 & 116  & 0.90 \\ 
\Simwhf & $20.5$ & $2.38$  & $0.84$ & $20.2$ &
$2.3\pm1.7$ & $8.0\pm2.1$ & 0.22 & 92.0  & 0.77 \\ 
\Simlf & $12.4$ & $3.75$  & $1.38$ & $20.5$ &
$0.9\pm0.5$  & $3.3\pm0.7$ & 0.13  & 26.5 & 1.41 \\
\Simhf\, $250\kyr$ & $10.9$ & $6.73$  & $2.80$  & $20.8$ &
$2.4\pm3.3$ &  $6.3\pm3.3$ & 0.09 & 108 & 3.22 \\
\Simhf & $10.9$ & $7.23$  & $2.72$  & $20.8$ &
$2.3\pm1.3$ &  $8.2\pm2.2$ & 0.09 & 73.5 & 2.02 \\
\Simll\, $250\kyr$ & $3.1$ & $4.46$  & $ 1.82$  & $20.6$ &
$3.2\pm3.8$ &  $7.0\pm4.0$ & 0.06 & 38.7 & 1.74 \\
\Simll & $5.79$ & $2.80$  & $ 0.99$  & $20.3$ &
$2.0\pm0.9$ &  $6.1\pm1.9$ & 0.10 & 19.0 & 1.36 \\
\Simk4 & $2.78$ & $3.44$  & $1.16$  & $20.4$ &
$1.2\pm0.7$ &  $6.2\pm0.9$ & 0.06 & 42.2 & 2.45\\
\Simsb & $3.39$ & $2.76$  & $1.00$  & $20.3$ &
$1.6\pm0.8$ & $3.4\pm1.5$ & 0.08 & 38.2 & 2.77 \\
\Simbb & $51.1$ & $4.76$  & $1.94$ & $20.6$ &
$1.0\pm0.5$ & $2.9\pm0.7$ & 0.24& 42.5 & 1.78\\ 
\hline
\end{tabular}
\caption{Results of the parameter study at $t=500\kyr$. Listed are the mass, mean
  density, mean surface density, corresponding column depth, velocity dispersion  and the
  x- velocity away from the source of the most prominent
  structure. Then the mean diameter (see Eq. \ref{diameter}) of the
  pillar and the mean density of the hot gas are given. Finally, the
  pressure difference $\Delta P$ (see Eq. \ref{delta_P}) is
  listed. For the fiducial simulation as well as the more rapidly
  evolving simulations these quantities are given at an earlier stage
  ($t=250\kyr$) as well.\label{TAB_compare}}
\end{center}
\end{table*}

For a quantitative discussion we investigate the most prominent pillar
structure in each
simulation in more detail. To define the tip we take the particle closest to the
source of radiation\footnote{In \Simll, \SimMh\, and \Simsb\, the second
  tip was taken to produce comparable results.} above a threshold density of
$\rho_\mathrm{tresh}=10^4\dens$. We then take its surrounding, the
region spanning $1\pc$ in the x-direction and 
$.3\pc$ in the negative and positive y- and z- direction. The cold,
un-ionized gas in this region is defined as the
pillar\footnote{Since we only take the unionized gas inside the
  surrounding the actual volume of the pillar changes from simulation
  to simulation.}. This definition allows us to extract the important 
quantities of the most prominent structure by the same algorithm for
all simulations. The characteristic values, which allow for a
comparison with the observations as well as a deduction of the
underlying physics, are given in Table 
\ref{TAB_compare}  for the defined pillar at
$t_\mathrm{final}=500\kyr$. Denote that due to the adaptive nature of
SPH we have a very high resolution in the prominent structures. The
pillar in \Simfid\, e.g. contains $5.6\times10^4$ particles, which
corresponds roughly to a spatial resolution of $178\times17.8\times17.8$ on the
$1\pc\times0.1\pc\times0.1\pc$ of this pillar. This resolution is up to
now unprecedented and allows for a detailed comparison of the kinematics
of this pillars with the observations (see \S \ref{Comp_Obs}). 

We calculate the diameter of a pillar via 
\begin{equation}
\label{diameter}
d_\mathrm{pillar}=2\sqrt{\frac{M}{\rho x \pi}}
\end{equation}
with $x=1\pc$ as the length of the pillar. In addition, 
the mean density\footnote{Denote that we 
  always give the real density $\rho$ of the hot gas, not the number
  density $n$ to avoid the factor of $\mu = 0.5$ when comparing the low density,
  un-ionized gas to the ionized gas.} of the
hot gas is listed in Table \ref{TAB_compare}.  
It is notable that all simulations with
the same impinging flux share a similar density of the pillar as well
as of the hot gas. 

As both gas phases are treated isothermal
(cf Eq. \ref{P_iso}) the pressure difference at $t_\mathrm{final}=500\kyr$ is:
\begin{equation}
\label{delta_P}
\Delta P_\mathrm{final} = \frac{P_\mathrm{ion,final}}{P_\mathrm{pillar,final}} =
\frac{2T_\mathrm{ion}\,\rho_\mathrm{ion,final}}
{T_\mathrm{nion}\,\rho_\mathrm{pillar,final}}.
\end{equation}
$\Delta P_\mathrm{final}$ is very close to
unity for all simulations \footnote{In fact,
  the value is always slightly above one, but this can be
  e.g. attributed to the complete neglection of the turbulent motion
  in the cold gas in Eq. \ref{delta_P}.}. Therefore, we derive as a first result that the pillars are in
thermal equilibrium with the hot surrounding gas.

\subsection{Resolution and Boundary Conditions}
\label{RES_res}
The first two simulations do not address different physical properties
but rather numerical details. Simulation \Simlr\, was performed with exactly the
same setup as \Simfid, but with eight times less particles. This leads
to a two times lower spatial resolution. Nevertheless, the
morphology is comparable to the high resolution case (Fig
\ref{FIG_compare}, panel 2). The only noticeable
difference is that the second largest structure in this case has
already merged with the third structure. Furthermore, tiny
structures are less frequent. The physical properties
(see Table \ref{TAB_compare}) are similar as well. The
structures in the low resolution case tend to be a bit more massive,
which can be expected. Altogether, the morphology and the global
physical properties are comparable and thus, we conclude that
\Simfid\, is reasonably converged.

In the other test case we investigate the boundary condition in the negative
x-direction. In \Simopen\, this boundary is not reflecting. Instead the
gas is allowed to stream away freely. This leads subsequently to a lower
density in the ionized region. As a consequence \Simopen\,
(Fig. \ref{FIG_compare}, panel 3) looks similar to \Simhf\, (panel 8, see \S
\ref{RES_flux}), as the radiation is 
able to penetrate further into the computational domain. Nevertheless,
the formation of pillars still 
takes place and is not strongly affected. Even density and mass
assembly of the most prominent structure are alike. Therefore,
the choice of the boundary condition does not influence the overall
scenario significantly. As it is more realistic to assume hot gas is
already present in the region between the ionizing source and the
simulated part of the molecular cloud 
we keep the reflecting boundary condition in all other
simulations. These reflection can be interpreted as flux conservation at the
left border of the simulation: as much gas streams from the area
towards the source into the region as is streaming outwards.

\subsection{Turbulent Mach Number}
\label{RES_Mach}
A main purpose of this study is to disentangle the effects of the
initial turbulent density distribution on the formation of the
pillars.
Therefore, the level of turbulence is changed. We 
take different turbulent setups: One has been evolved from a
very high Mach number (Mach 20) to \SimMvh. The other four represent
different stages of the decay starting from our fiducial turbulent
setup (G09b) at Mach 10. They are taken at \SimMh, \Simfid, \SimMl\,
and \SimMvl, respectively. When
non-driven turbulence decays, most power is lost on the large scale
modes. This can be seen in Fig. \ref{FIG_evol}: In \Simfid\,(column 2)
and \SimMh\, (column 3) the surface density is
clearly dominated by the large modes, which form the prominent
fingers. In contrast, in \SimMvl\, (column 1) no significant
pillars evolve, since the initial density distribution is already too
smooth and the dominant mode has decayed to far. This trend can already be
seen in \SimMl\, (Fig. \ref{FIG_compare}, panel 4), where the
structures are less distinct.
\SimMvh\, (column 4) is a much more violent case. Since there is a lot
of power on the largest density scale, structures are evolving. However, these 
are already being torn apart at the same time, as discussed in \S
\ref{Globules}. 

Overall, the evolution is mainly dominated
by the pressure differences between hot and cold gas. Compared to the
increase in the pressure due to the ionization (three orders of magnitude)
the differences from varying the Mach number are 
small. However, a small trend is visible in the average
density of the assembled structure (see Table \ref{TAB_compare}). The
higher the Mach number, the
higher the density of the formed structure. That can be directly
related to the density of the initial turbulent filament, which is
<denser at a higher Mach number. This effect is also visible
from the first row of Fig. \ref{FIG_evol}, where the absolute value in
the area of highest density is increasing from left to right. 

\subsection{Temperature and Pressure}
\label{RES_T}
The most striking difference can be seen between
\Simfid\,(Fig. \ref{FIG_compare}, panel 1) and \Simwif\,(panel 5). Both
initial conditions are self-similar, both were set up with the same
initial random seed for the turbulence and they are relaxed until
their velocities resemble Mach 5 at their respective
temperature. Consequently, at the time the ionization is switched on,
their density distribution is identical. Since the impinging flux is
the same, the radiation ionizes the same regions in both
cases. Nevertheless, in \Simwif\, no evolution of any filamentary
structure is visible. This leads to the conclusion that the pressure
balance between the hot, ionized and the cold, un-ionized gas plays a
crucial role in the formation of structures.

Taking Eq. \ref{P_iso} and the straightforward assumption that only
regions with a pressure lower than the pressure of the hot gas can
be compressed gives 
\begin{equation}
\label{P_crit}
P_\mathrm{nion,initial} \le P_\mathrm{ion,initial}
\end{equation}
and thus
\begin{equation}
\rho_\mathrm{nion} \le \rho_\mathrm{ion}\frac{2T_\mathrm{ion}}{T_\mathrm{nion}}.
\end{equation}
If we assume $\rho_\mathrm{ion}=100\dens$ in the beginning, as the ionization
mainly penetrates the lower density regions, this equation yields
$\rho_\mathrm{nion,10K} \le 3.6\times 10^{5}\dens$ and
$\rho_\mathrm{nion,100K} \le 3.6\times 10^{4}\dens$. The maximum density
$\rho_\mathrm{max}=8.8\times 10^{4}\dens$ 
in both simulations lies in between these thresholds. Thus, in
\Simfid\, the
pressure of the ionized gas is high enough to compress even the
densest structures, whereas in \Simwif\, several regions are able
to resist the compression. Therefore, the ionized 'valleys' are not
expanding significantly in the tangential direction, the density is
not lowered as much and the ionization is not able to reach much
further. At the same time the un-ionized 'hills' are less compressed,
but since they are closer to the front they are accelerated more
strongly in the x-direction and the initial differences in the front
position are leveled out.

In general, the formation of
pillars critically depends on whether the density contrast between
the dense regions ($\rho_\mathrm{high}$), which can not be ionized,
and the less dense regions ($\rho_\mathrm{low}$), which can be
ionized, is lower than the temperature ratio between ionized and
un-ionized gas. By defining the density 
contrast in the initial conditions as $\Delta \rho_\mathrm{init} =
\rho_\mathrm{high} / \rho_\mathrm{low}$ and taking
into account Eq. \ref{P_crit} the critical criterion for the formation
of pillars can be written as:
\begin{equation}
\label{delta_rho}
\Delta \rho_\mathrm{init} \le \frac{2T_\mathrm{ion}}{T_\mathrm{nion}}.
\end{equation}
Since stars will only form in compressed regions, e.g. pillars, this gives
an estimate if a region will undergo triggered star formation.

To test this condition we used the same warm initial conditions but
increased the ionizing flux in \Simwhf. The flux is now at the level
of \Simhf\, (see \S \ref{RES_flux}). This is equivalent to increasing
the lowest density which can still be ionized, i.e. to increasing
$\rho_\mathrm{low}$. As a result, $\Delta \rho_\mathrm{init}$ is decreased
and thus structure formation is possible again in the warm case
according to Eq. \ref{delta_rho}. In fact, it can be seen in
Fig. \ref{FIG_compare} (panel 6) that indeed pillar formation is triggered again.

Unfortunately there is no straightforward way to define the density
contrast in a turbulent medium, especially since the impinging flux
plays a major role in defining 'high' and 'low' density as seen in
\Simwhf. However, Eq. \ref{delta_rho} already shows, that increasing
the mean density $\bar{\rho}$ while keeping the temperature constant
will not help to hinder the formation of pillars. This is supported by
the fact that pillars form in \Simfid\, as well as in \Simll\, (see \S
\ref{RES_flux}). 

\subsection{Initial Flux and Density}
\label{RES_flux}
In the next test we vary the impinging photon flux. 
As in the simulations performed in \S \ref{method},
the flux is able to ionize immediately $0.55\%$, $1.67\%$ and $5\%$ of a
medium at a constant density of $\rho=300\dens$  in \Simlf,
\Simfid\,(intermediate flux)
and \Simhf, respectively.
\begin{figure}
  \centering{
   \plotone{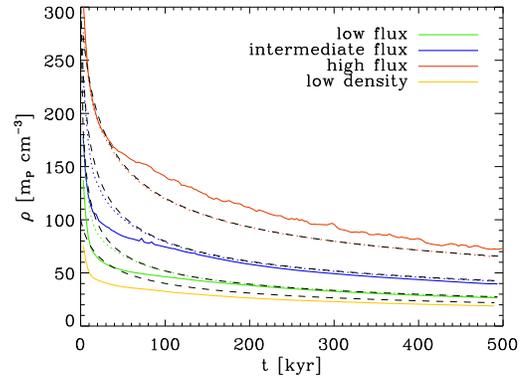}
  \caption{The change of the mean density in the hot gas during the
    simulated time. Solid lines: green \Simlf, blue \Simfid, red
    \Simhf\, and yellow \Simll. Dotted lines: the comparison simulations of \S
    \ref{method} at the respective flux. Dashed black lines: the
    analytical solution according to Eq. \ref{rho_Stroem} with
    $f=\sqrt{\frac{5}{4}}$ \label{rho_evol}}}
\end{figure}
The evolution of the density in the hot component is shown in
Fig. \ref{rho_evol}. Although the medium is highly turbulent
Eq. \ref{rho_front} still gives a very good estimate of this density,
especially after an initial phase. Only the case with the high flux
differs from the analytical solution. This can be understood as  
Eq. \ref{rho_Stroem} depends on both the penetration depth and the
density of the ionized gas. A higher flux can ionize a larger fraction
of the computational volume straight away. Equivalently, the evolution
of high flux would follow Eq. 4 more closely if the mean gas density
would be higher, thus resulting in a shorter penetration length. As
long as the penetration length is relatively small ($<5\%$), the
turbulent case is still comparable with the case of a constant flux
and Eq. \ref{rho_front} still represents a valid description of the
evolution of a turbulent HII region. 

From a morphological point of view, the pillars in the simulations with a
higher flux (i.e. when the computational volume is located closer to the
ionizing source) are smaller than in the case of a lower flux
(Fig. \ref{FIG_compare}, panel 8, panel 1 and panel 7 in decreasing
flux order). In addition, they gain more momentum away from  
the source and move faster away from the source as the
photo-evaporation is stronger. At the same time the density of the hot
gas is higher, leading to denser, more compressed structures with a
smaller diameter. Due to the higher photo-evaporation rate their
average masses are as well lower (see Table \ref{TAB_compare}).

Changing the initial flux is expected to have a similar effect as
changing the mean density, as $\rho\propto x^2$ (cf
Eq. \ref{x_Stroem}). In \Simll\, we reduced the 
mean density by a factor of three. At the same time we reduced the
flux by a factor of three to avoid an extremely high level of ionization
degree. In total, this corresponds to the same penetration length as
in \Simhf. Thus, we expect a similar morphology to evolve, but the
densities should be lower. In
Fig. \ref{FIG_compare} (panel 9) this can be clearly seen. The
morphology is similar to \Simhf, the front is at a similar
position. Again, the density (Fig. \ref{rho_evol}) in the hot gas 
evolves similarly to the expectation for a homogeneous medium. The
mass assembled 
in the most prominent structure (Table \ref{TAB_compare}) is lower and the
density of the structure fits the findings of pressure equilibrium
(see \S \ref{RES_T}).

Combining these findings with the results of \S \ref{RES_general}
allows us to make an interesting prediction. As the density of the
hot gas behaves similarly to the case of a homogeneous medium and as the
structures are in approximate pressure equilibrium with the
surrounding hot gas, we can predict the 
density of the structures from the initial mean density of the medium,
the flux of the source, and the time since the ignition of the source
or the position of the ionization front.
The density of the forming structures is thus given as
\begin{equation}
\label{rho_pred}
\rho_\mathrm{pillar}\approx\frac{2T_\mathrm{ion}}{T_\mathrm{nion}}\rho_\mathrm{ion} \approx
\frac{2T_\mathrm{ion}}{T_\mathrm{nion}}\rho_0\left(1+\frac{5}{4}\frac{c_\mathrm{s,hot}}{x_\mathrm{s}}(t-t_0)\right)^{-\frac{4}{10}},
\end{equation}
where we used Eq. \ref{rho_front}. $x_\mathrm{s}$ depends on the
initial density and the impinging flux (see Eq. \ref{x_Stroem}). As we
expect the assumption of pressure equilibrium to hold in the case of a
point source, the three-dimensional expression taking into account
geometrical dilution is given by 
\begin{equation}
\rho_\mathrm{pillar}\approx
\frac{2T_\mathrm{ion}}{T_\mathrm{nion}}\rho_0\left(1+\frac{7}{4}\frac{c_\mathrm{s,hot}}{x_\mathrm{s}}(t-t_0)\right)^{-\frac{2}{7}},
\end{equation}
where $R_\mathrm{S}$ is the Str{\"o}mgren radius (for a detailed
derivation of the three-dimensional front position see e.g. \citealt{1991pagd.book.....S}).

\subsection{Turbulent Scale}
\label{RES_scale}
\begin{figure}
  \centering{
   \plotone{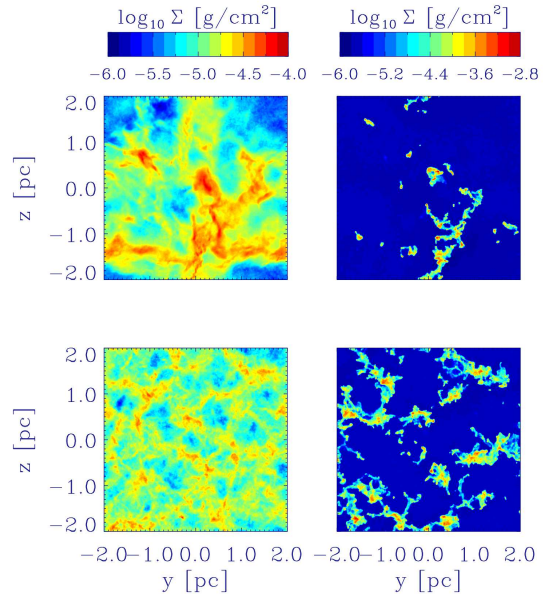}
  \caption{Projected surface density along the x-axis. The projected
    slice is always $2\pc$ thick. Left: $t=0\kyr$ and
    the slice starts at the surface facing the O-star,
    right: $500\kyr$ and the slice is adjusted to encompass the
    substructures. Top row: only modes $k=1-4$ are populated
    initially, bottom row: only modes $k=4-8$ are populated
    initially.\label{turb_surf}}}
\end{figure}

To study the effect of the largest scales of  the initial turbulence
(the turbulent input scales) we compare \Simfid\, with \Simk4, a run 
in which we populate modes $k=4...8$, instead of $k=1..4$ as
usual. The resulting surface density in
the first $2\pc$ facing the star is shown in
Fig. \ref{turb_surf}. Already in the initial conditions (left column) a
clear difference can be seen. Whereas the power on the larger k modes
leads to large, distinct structures (top panel), power on the smaller modes show
already a much more diversified density distribution (lower
panel). After $t_\mathrm{final}=500\kyr$ (right column) the ionization
leads to an enhancement of the pre-existing structure. The densest filaments
survive, while the other material is swept away by the ionization. In
\Simfid\, (top panel) this leads to an excavation of the few, but
bigger structures and thus to the creation of few, but distinct
pillars. On the other hand, in \Simk4\,(bottom panel) more structures,
but of smaller scales survive, which leads to several, but more diffuse structures.

Together with \S\ref{RES_Mach} this shows, that only a strong enough
turbulent driving on a large enough driving scale
leads to the formation of coherent structures as seen in observations. 
As has been shown in \S \ref{RES_res} this is not an
effect of the resolution. The turbulence is well enough resolved to
allow for small enough modes to produce fuzzy structure in \Simfid,
but the evolution under the influence of UV-radiation is dominated by
the larger modes.

Another possibility to change the input scale of the turbulence is to
simply increase or decrease the size of the simulation domain. In
\Simbb\, the box size is doubled to $8\pc$. Since the particle number is kept
constant, this leads to a factor of two lower spatial resolution. So
the resolution in the part of the domain shown in Fig \ref{FIG_compare} (panel
12) is comparable to the low resolution case \Simlr\,(panel 2). In \Simsb\,
the box size is halved to $2\pc$.
This corresponds to doubling the spatial resolution. The domain
has a smaller extent in the x-direction as well, which we
compensate by taking two times the evolved turbulent box in the
x-direction. This is valid, since the initial conditions were evolved
with periodic boundary conditions. The particle number is thus
$4\times10^6$, twice as high as in most other cases. \Simsb\,
is the situation in between \Simfid\,and \Simk4, as the
largest mode is $2\pc$, which corresponds to a $k=2$ mode in a $4\pc$
domain.

As the density distribution and therefore the density contrast is
self-similar in all three cases we expect that the same regions of the
initial conditions will form the dominant structures. Only the size of
the encompassed region and therefore the size and mass of the pillars
should change, if the process is indeed scale free. In
Fig. \ref{FIG_compare} all three simulations (panel 
1, panel 11, panel 12) show a clear sign of the largest $k=1$
mode. The size of the structures formed is linearly dependent
on the initial box-length or size of the largest k-mode. The values of
Table \ref{TAB_compare} confirm the importance 
of the largest mode. In the assembled structures the estimated
diameters are roughly a factor of two different and the masses vary by
a factor of four. This is as expected since the regions initially
encompassed by the radiation should differ by a factor of two in the
y- and the z-direction.

Taking these results on the turbulent scale into account, we conclude
that the mass and size of the pillars is directly dependent on the input
scale of the turbulence, e.g the size of the driving process or the
size of the pre-existing molecular cloud. On average, the most
prominent structures in our simulations with an intermediate flux are
$d_\mathrm{pil}\approx\frac{1}{40}x_\mathrm{turb}$, where
$x_\mathrm{turb}$ is the largest turbulent input mode\footnote{In fact
the value for \Simll\, from Table \ref{TAB_compare} does not match precisely,
but from Fig. \ref{FIG_compare} the factor of $\approx4$ between
\Simfid\, and \Simll\, can be seen.}.

\subsection{Star Formation}
\label{RES_SF}
In several simulations triggered dense regions form cores and are
driven into gravitational collapse. Since
star formation is not the main goal of this study we do not replace
them by sink particles. Instead we remove the particles forming a core
from the simulation to avoid a considerable slowdown of the
calculation. Following G09a we define a core as all gas with a density
above $\rho_\mathrm{crit}=10^7\dens$ in the region around the density
peak and remove the particles representing this core. 
The core formation is not a numerical effect, since the resolution limit
as given by \cite{1997MNRAS.288.1060B} is $\rho_\mathrm{num}=3\times
10^8\dens$ in the lowest resolution 
case. We give the simulation, mass\footnote{The masses do not differ 
  significantly, as we do not follow the further accretion process and
  at the moment of formation the cores are still similar.},
formation time, average speed 
away from the source and positions in Table \ref{star_form}. If we
assume the cores 
to be decoupled from the rest of the cold gas then their position at
the end of the simulation can be estimated by these position and
velocities. All of them are still close to the prominent structures,
some are traveling further inside the structures, some are
lagging behind and would by now be slightly outside of the pillar,
closer to the source.

\begin{table}
\begin{center}
\begin{tabular}{lcccccc}
\hline
Simulation & $t_\mathrm{form}$ & $M$ & $v_\mathrm{x}$ & $x$ & $y$ & $z$ \\
 & $t[\kyr]$ & $[\Msun]$ & $[\kms]$ & $[\pc]$ & $[\pc]$ & $[\pc]$ \\
\hline
\Simbb &  305.3 & 0.86 & $4.37\pm1.4$ & -2.83 & -1.84 & 1.26 \\
\Simbb &  353.3 & 0.90 & $4.16\pm0.7$ & -2.60 & -1.86 & 1.34  \\
\Simbb &  403.7 & 0.78 & $4.53\pm1.1$ & -2.58 &  0.29 & 0.41  \\
\Simbb &  469.3 & 0.83 & $4.99\pm0.7$ & -2.06 & -1.88 & 1.48  \\
\Simhf &  429.5 & 0.57 & $11.27\pm1.0$ & 1.86 & -1.05 & 0.64  \\
\Simfid &  493.3 & 0.72 & $3.92\pm0.8$ & 0.19 & 0.48 & -0.79  \\
\hline
\end{tabular}
\caption{Listing of the proto-stellar cores forming in the different
  simulations. Given are mass, formation time, average speed away from
  the source and their formation position.\label{star_form}} 
\end{center}
\end{table}

\begin{figure}
\plotone{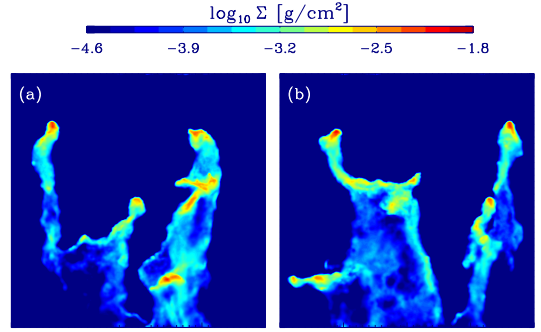}
 \caption{Projected surface density of the pillars in the simulation
    with Mach $7$. {\bf (a)} x-y projection, {\bf (b)} x-z
    projection. Colour-coded is the surface density, each 
    figure is $1\pc \times 1\pc$. The density enhancements inside the
    pillars at the left hand side correspond directly to caps, which
    are not shadowed by the leading tip on the right hand side and
    vice versa. \label{fig_M7}}
\end{figure}

\begin{table}
\begin{center}
\begin{tabular}{|c|ccc|ccc|}
\hline
 Simulation & & Mach 5 & & & Mach 7 & \\
 & Tip 1 & Tip 2 & Tip 3  & Tip 1 & Tip 2 & Tip 3 \\
\hline
 M $[\Msun]$  & 0.62 & 1.87 & 0.35 & 0.79 & 0.45 & 0.48 \\
 $v^{\mathrm{T}}_0 [\kms]$ & 0.60 & 1.08& 0.54& 0.98 & 0.56 & 1.04 \\
 $v^{\mathrm{T}}_{500} [\kms]$  & 0.10 & 0.47 & 0.25 & 1.24 & 0.32 & 0.45 \\
\hline

Simulation & & Mach 7 & & & Mach 12.5 & \\
 & Glob. 1 & Glob. 2 &  Glob. 3 & Glob. 1 & Tip 1 &  Tip 2 \\
\hline
 M $[\Msun]$  & 0.32 & 0.15 & 0.20 & 0.12 & 1.12 & 0.54 \\
$v^{\mathrm{T}}_0 [\kms]$ & 2.44 & 2.31 & 2.20 & 3.98 &
1.36 & 0.55 \\
 $v^{\mathrm{T}}_{500} [\kms]$  & 3.00 & 2.85 & 2.05 &
 2.19 & 2.42 & 0.78 \\
 \hline
\end{tabular}
\end{center}
\caption{Properties of the globules and tips in the four
  simulations. The structures are chosen from the final stage at
  $t=500\kyr$ , the numbers are given from top to
  bottom. Listed are the mass, the initial 
  tangential velocity ($v^{\mathrm{T}}_0$) of the particles that are
  going to form the structure later, as well as their final tangential
  velocity ($v^{\mathrm{T}}_{500}$). \label{tab_globule}}
\end{table}

As it can be expected, in \Simbb, where the compressed  structures are
most massive, star formation is most frequent and happens earliest. 
There is an age spread present as well - the earlier a core forms the
closer to the source it will be. 
The other simulations where cores form
during the first $500\kyr$ after the ignition of the O star are \Simhf\,
and \Simfid. The higher flux leads to a higher compression
and thus an earlier formation of a core in \Simhf. due to the
higher photo-evaporation rate this core is also less massive and
moves faster compared to the core in \Simfid.

Altogether, triggered star formation is very likely in this
scenario. The cores form at the center of the structures, but since
their velocities differ from the velocity of their parental structure
they can be decoupled and either wander further into the trunk or lag
behind. This depends on the specific environment and does show no
 correlation with e.g. the initial flux. At the time they become
observable there might be no clear correlation to their birthplace any
more.

\section{Pillars, Caps and Globules}
\label{Globules}
Another interesting feature can be found when looking at different
projections of the simulation \SimMh\, (Fig. \ref{fig_M7}). By
comparing both surface densities it becomes clear that the density
enhancements inside the pillars are caps, which are exposed directly
to the UV-radiation in the
other projection. Thus, the small steps and wiggles seen in the
observations can be explained
directly by the sweeping up of smaller caps, which are not shadowed by
the leading tips of the pillars.

As a last point it is interesting to examine the stability of the
forming structures. In Table \ref{tab_globule} we show the correlation
between the formation of globules or pillars and the initial turbulent
velocity perpendicular to the plane of irradiation of the particles
forming these structures. Here we determine the mass of a globule by integrating over its material with a density higher than $n_{\mathrm{c}}=10^4\ndens$. A tip is defined
similarly by taking all the material above $n_{\mathrm{c}}$ in a sphere
of $R_{\mathrm{c}}=0.025\pc$ around the local density maximum.  
Table \ref{tab_globule} lists the most important features seen in the
final stage of Fig. \ref{FIG_evol}. In the physical properties
of the tips and globules a striking correlation can be seen. The tips 
have lower initial and final tangential velocities, the globules
much higher ones. Thus, a pillar can only be assembled when the leading blob is
moving slow enough perpendicular to the irradiation that a
sub-structure can survive in its shadow. Therefore, an
environment with lower Mach number favors the formation of stable,
massive structures, like in the simulation with \Simfid. \SimMh\,
represents the intermediate case where tips and globules form. The
case of \SimMvh\, corresponds to a rather violent scenario. Here, globule 1
has already decoupled from the rest of the gas. In addition, even the
most prominent pillar (Tip 1) has a high velocity and represents a
transient stage, which soon decouples from its stem and gets
disrupted during the next $100\kyr$. In
contrast, Tip 2 is an exceptional case - due to its very low tangential
velocity it allows for a stable pillar even in this violent
environment\footnote{Both statements have been verified by
continuing this simulation until $t=600\kyr$.}.

Furthermore, it is worth noting that the tangential velocity is not
affected strongly by the UV-radiation. This is reasonable, since the
surviving structures experience tangential compression from all sides,
so the net effect of the ionization balances out.
In general, our simulations are able to explain the formation of the
observed low mass globules or globulettes. Especially the similarity
between the structures forming in \SimMh\, and the structures observed in
the Rosette Nebula (\citealt{Gahm:2007lr}, their Fig. 3) is striking.

\section{Comparison to Observations}
\label{Comp_Obs}
\subsection{General Properties}
At first, we compare the density of the hot gas to observations. 
\cite{2002ApJ...581..335L} estimate the electron
density of the HII-region in the Trifid Nebula\footnote{The exciting
  source of the Trifid Nebula is HD 164492A an O7V star
  \citep{Lynds:1985qy}, so the UV-flux is comparable to our simulations.} from OIII as
$n_\mathrm{e}=50\ndens$. In a fully ionized region this corresponds
directly to the density given in Table 
\ref{TAB_compare}, since $n_\mathrm{e}=n_\mathrm{H}=\rho/m_P$. Thus,
the observed value is very similar to the density found in all
simulations with an intermediate flux at $t=500\kyr$. 
The average density of the pillars is around $10^4\dens$,
depending on the individual simulation. This is in very good agreement
with recent results from Herschel by \citet{Schneider:2010fk}, where
they find a typical average density of  $1.1\times 10^4\dens$ in the
cold structures in the Rosette Nebula\footnote{The structures are in
  the 'Extended Ridge' which is roughly $10\pc$ away from the several
  O-stars (O4 to O9) of NGC 2244. Thus, the situation there is as well
comparable to our simulations.}. In the Trifid Nebula,
\cite{2002ApJ...581..335L} estimate
$N(\mathrm{CS})\approx1.8\times10^{13}\sdens$ for the column density
of the dense core, which corresponds to 
$\mathrm{log}_{10}[N(\mathrm{H_2})/\sdens]\approx22.5$ with their
conversion factor. 
To compute the H2 column density we apply a conversion factor of
$\chi=N(\mathrm{H_2})/\Sigma=0.35$, using a hydrogen abundance of
$X=0.7$ and assuming that all hydrogen is molecular at these
densities. In all cases where structures form the column densities are
around $\mathrm{log}_{10}[N(\mathrm{H_2})\sdens]\approx20.5$ (Table
\ref{TAB_compare}).
As this is the averaged surface density of
the entire structure in our simulations it is two orders of magnitude lower than the
observed values for the dense cores. In the tips of the
pillars (see e.g. Fig. \ref{fig_M5}) the peak surface density is
$\mathrm{log}_{10}[\Sigma_\mathrm{max}/(\mathrm{g}\thinspace\mathrm{cm}^{-2})]=-1.2$
which leads to a column density $\mathrm{log}_{10}[N(\mathrm{H_2})/\sdens]=22.12$, which
is in good agreement with observations. 
\citet{2002ApJ...570..749T} estimate
$\mathrm{log}_{10}[N(\mathrm{H_2})/\sdens]\approx21.3$ for the most prominent
of the pillars in M16.
In RCW 120
\citep[e.g.][]{2009A&A...496..177D} condensation 4 seems to be a good
candidate for triggered star formation. The peak surface 
density is $\mathrm{log}_{10}[N(\mathrm{H_2})/\sdens]=22.15-22.49$. In addition,
\cite{2009arXiv0902.4751U} investigate a sample of 60 bright rimmed clouds
and find the column densities in cases of triggered star formation (that
is in the cases with photodissociation regions (PDRs)) to be
$\mathrm{log}_{10}[N(\mathrm{H_2})/\sdens]=20.9-22.8$. 

As all these observations match our simulations, we
conclude that the evolution of the density of the hot gas is
in good approximation given by the estimate in the case of a
homogeneous medium (Eq. \ref{rho_front}). Furthermore, since the
densities of the 
compressed structures are reproduced as well, we can assume that the
structures are indeed in pressure equilibrium with the hot,
surrounding gas. This provides the opportunity to determine
the density of the hot gas and the compressed structures directly
from the initial mean density, the flux from the source and the time
since the ignition of the source or the position of the ionization
front, respectively.

\subsection{Velocity Field of a Singular Pillar}
\label{OBS_vel}

\begin{figure}
 \plotone{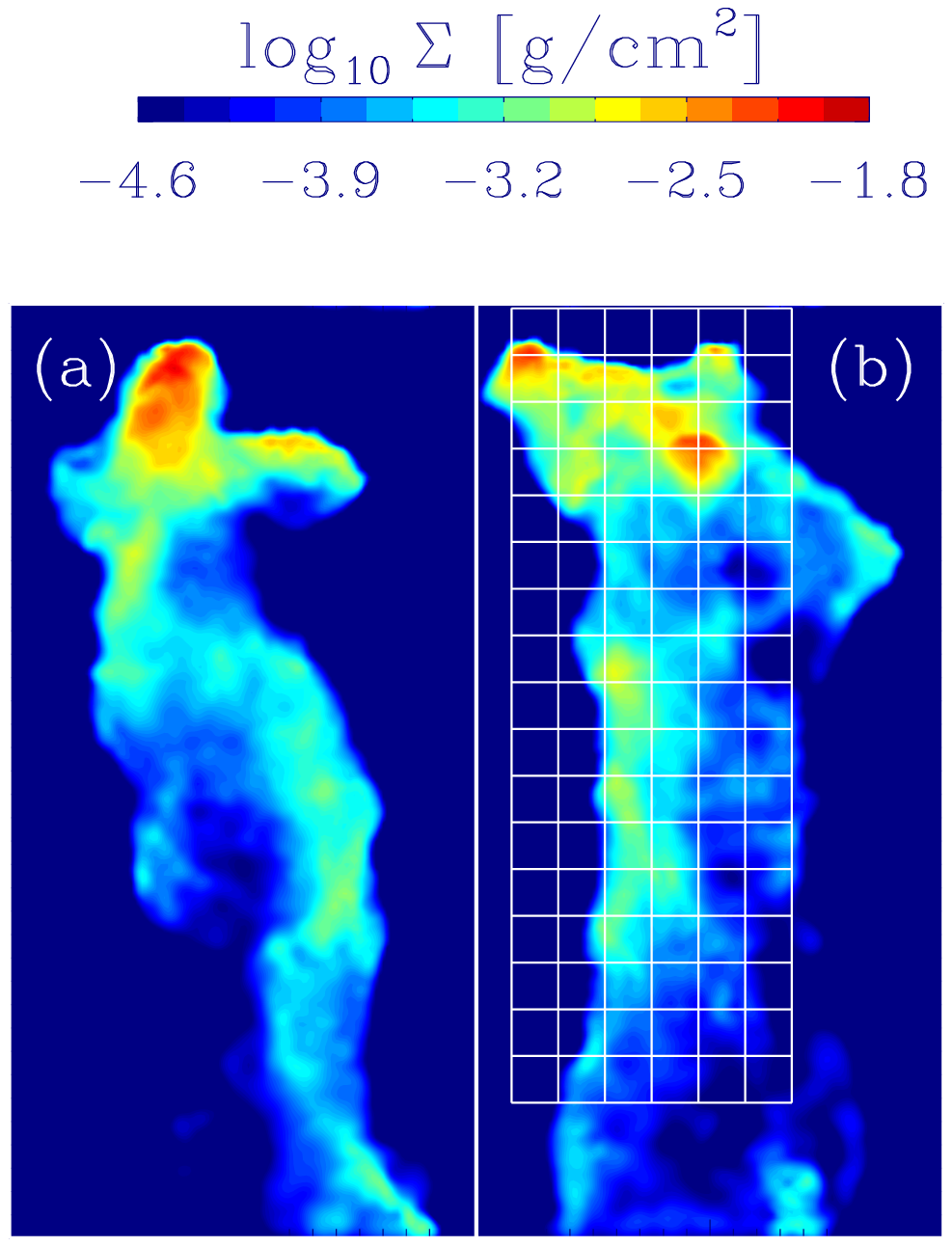}
  \caption{The most prominent pillar of the simulation with Mach
    5 in its reference frame. Colour-coded is the surface density, each 
    figure is $0.4\pc \times 0.8\pc$.  {\bf (a)} x-y projection, {\bf (b)} x-z
    projection. The superimposed 
    grid denotes the bins along which the line-of-sight (LOS)
    velocities in Fig. \ref{fig_M5los} are taken. In order to match
    the observational beam-size \citet{2006A&A...454..201G} each bin is
    $0.04\pc\times 0.04\pc$.\label{fig_M5}}
\plotone{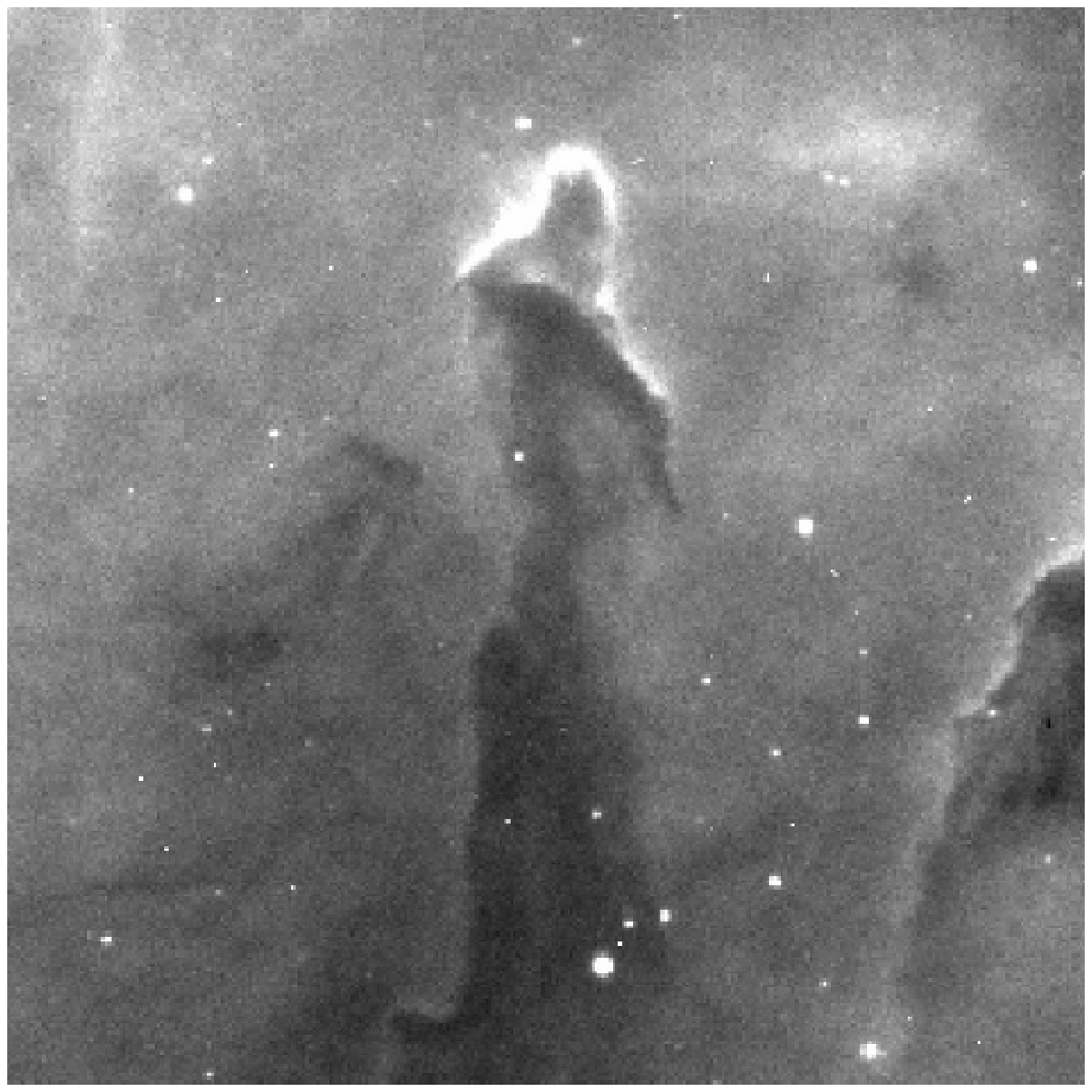}
  \caption{For a better comparison, the Dancing Queen (DQ)  trunk in NGC 7822
    as observed by \citet{2006A&A...454..201G} is shown. Depicted is a $0.8\pc \times 0.8\pc$
    subset of their Fig. 1, roughly to scale with our
    Fig. \ref{fig_M5}. \label{DQ}}
\end{figure}

\begin{figure}
\plotone{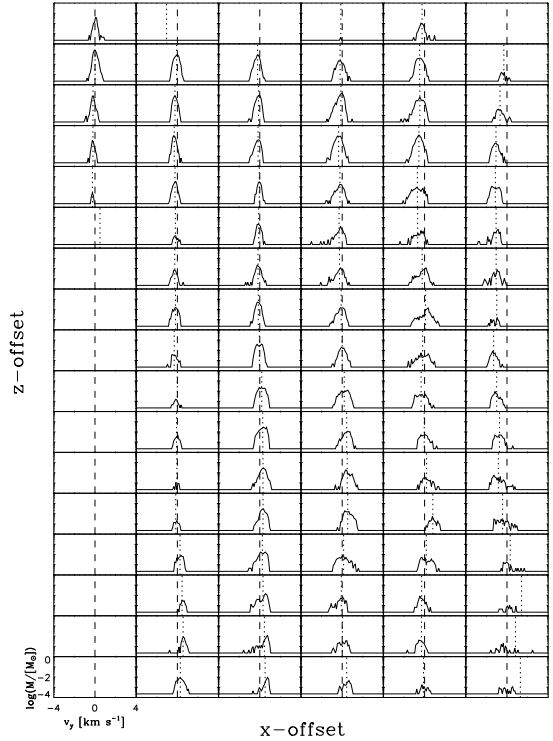}
 \caption{Line-of-sight(LOS) velocities along the most prominent
   pillar in the simulation with Mach 5. The LOS-velocities are taken in bins
    of the size $0.04\pc\times0.04\pc$ (see Fig. \ref{fig_M5}) to
    match the observational resolution. 
    The mean velocity in each bin is depicted by the dotted lines. In
    Fig. \ref{vy_vz} this velocities are plotted for column 2, 3 and
    4. Denote that in addition to the overall pattern, the line-width
    is in very good agreement with the
    observations \citet{2006A&A...454..201G}. \label{fig_M5los}}
\end{figure}

\begin{figure}
 \plotone{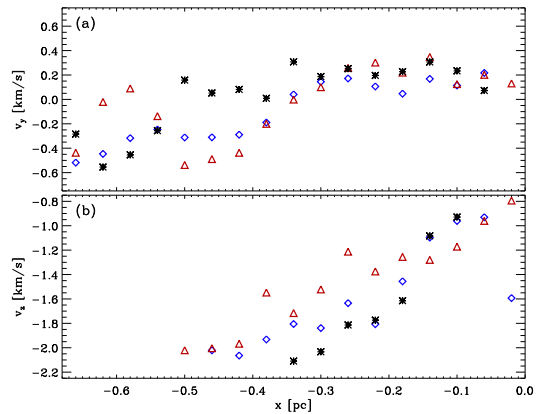}
\caption{The line-of-sight(LOS) velocity along the pillar.
    We show the LOS velocity of the most prominent
    pillar in the simulation with Mach 5 as function of position along
    three longitudinal cuts. The x-axis is parallel to the major axis.
    Stars, diamonds and triangles correspond to cuts parallel to the x-direction
    through the center (diamond) and the left (star) and right (triangle) side.
    For more details, see Figs. \ref{fig_M5} and \ref{fig_M5los}. The location of the head is at $x=0\pc$.
    Top panel {\bf (a)}: projection along the y-axis. Bottom {\bf (b)}: projection along the z-axis. A
    velocity gradient along the pillar is clearly visible. The gradient 
    matches the observations of the Dancing Queen Trunk very
    well \citet{2006A&A...454..201G}. The fact that the velocities for
    the different cuts rarely cross, especially in the 
    bottom plot, is a signature of rigid rotation. This, in
    combination with the overall gradient produces the so called
    cork-screw pattern, as observed.\label{vy_vz}}
\end{figure}

Another property to compare between simulations and
observations are the details of the velocity distribution in the pillars. Since
turbulence is a highly complex process it is hard to
precisely predict the outcome of simulations. Therefore, we do not
set up a simulation to match some specific observation but instead
start with realistic initial conditions and then look for an observed counterpart of
the outcome of our simulation. 
In the following we analyze the velocity structure of a typical pillar
in the Mach 5 simulation which has a similar morphology and mass as
the only trunk with well observed line-of-sight velocities,  the
Dancing Queen (DQ)  trunk in NGC 7822 (\citealt{2006A&A...454..201G},
see Fig. \ref{DQ}).
The authors give the diameter as $d_\mathrm{obs}\approx0.12-0.15\pc$, the total
estimated mass from $^{12}$CO is $M_\mathrm{obs}\approx9.2\Msun$. This 
is similar the simulated pillar ($d_\mathrm{sim}\approx0.12\pc$,
$M_\mathrm{sim}\approx12.6\Msun$, see Table \ref{TAB_compare}). If we
subdivide the pillar into a head and a body the mass splits up into
$M_\mathrm{head,s}\approx7.2\Msun$ and $M_\mathrm{body,s}\approx5.3\Msun$
(compared to $M_\mathrm{head,o}\approx5.7\Msun$ and
$M_\mathrm{body,o}\approx3.5\Msun$  in the DQ trunk).

\begin{figure}
\plotone{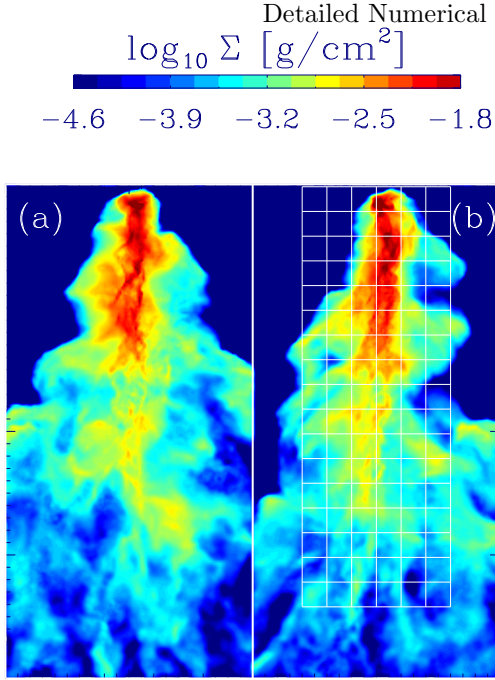}
 \caption{The main structure in the radiation driven implosion (RDI)
   model (G09a). Colour-coded is the surface density, each 
    figure is $0.4\pc \times 0.8\pc$.  {\bf (a)} x-y projection, {\bf (b)} x-z
    projection. The superimposed grid denotes the bins along which the
    LOS-velocities in Fig. \ref{fig_BElos} are taken. \label{fig_BE}}
\end{figure}

\begin{figure}
\plotone{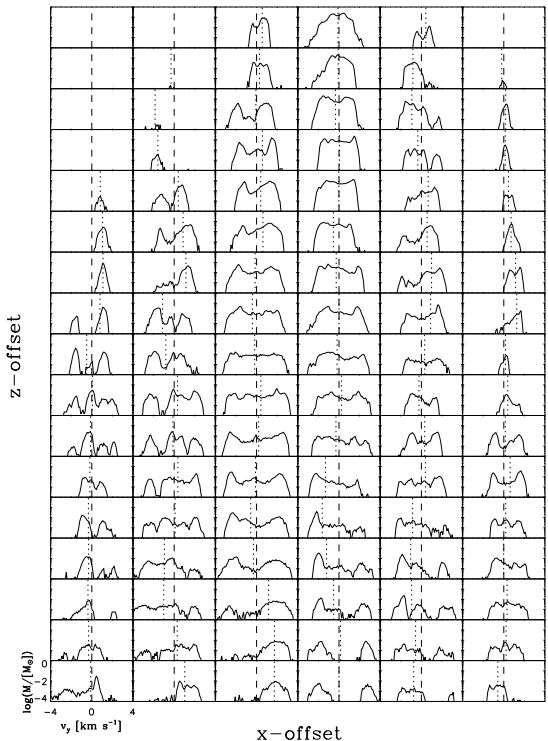}
 \caption{LOS-velocities along the main structure in the
   RDI-model (G09a). The LOS-velocities are taken in bins
    of the size $0.04\pc\times0.04\pc$ (see Fig. \ref{fig_BE}) to match the observational resolution.  Neither the overall pattern, nor the line-width are
   in agreement with the observations \citet{2006A&A...454..201G}. \label{fig_BElos}}
\end{figure}

To enable a more detailed comparison we impose a grid across the head
in the y-x plane (see Fig. \ref{fig_M5}) to
resemble the beams along which the line-of-sight (LOS) velocity was taken. We
divide the LOS-velocity, ranging from $v_\mathrm{z}=-4\kms$ to
$v_\mathrm{z}=4\kms$ into 80 equally sized bins. In each of the
velocity bins the mass is integrated. Fig. \ref{fig_M5los} shows the
profiles obtained in that way. As we do not take any
radiative transfer, temperature dependencies or chemistry into account
in our profiles,
they are not as smooth and symmetric as the observed HCO$^+$
profiles. Nevertheless, the 
similarities are striking. 

In our simulation the standard mean deviation for all lines with a
peak higher than $5\times 10^{-3}\Msun$ is on average $0.38\kms$, which
corresponds to a FWHM of $0.94\kms$. This is very similar to the
observed FWHM of about$1.2\kms$. Thus, the irregular motions in the
simulations and observations correspond very well. Our simulations are
missing the offset of $v_0\approx-16.4\kms$, the speed at which the
observed DQ is moving with respect to us. In addition, the profile is
reminiscent of a rotational pattern, as the peak is shifting from left
to right, 
as well as a gradient along the x-direction, since the peak is
shifting from top to bottom. This so called 'corkscrew' pattern (see
Fig. \ref{vy_vz}) has
often been attributed to magnetic fields.
However, our pure hydrodynamical simulations reproduce the pattern,
indicating that magnetic fields might play a minor role. Instead, the
pattern is produced by confining the motion of the turbulent cold gas
inside the pillars by the hot gas surrounding it.

As elongated, pillar-like structures can also be the result of the
ionization of pre-existing clumps (RDI-model, see
e.g. \citealt{Mackey:2010uq}), we applied the same 
analysis to our previous simulation of this scenario
(G09a). We took the case of a low ionizing flux
impinging onto a marginally stable Bonnor-Ebert-Sphere. The low flux
case was taken to allow for more moderate velocities. This scenario
results in an elongated feature depicted in Fig. \ref{fig_BE}
at $t=600\kyr$. The corresponding LOS-profiles are given in Fig.
\ref{fig_BElos}. Here, the profiles differ significantly from 
the observations as well as from Fig. \ref{fig_M5los}. First of all,
there is a double peak. This shows that the structure is produced by
the collision of two shock-fronts, one moving away from us and one moving
towards us. These two shock fronts, which are encompassing the original density
enhancements, can be seen directly colliding in the perpendicular
projection (Fig. \ref{fig_BE}a).

Even if the two peaks become indistinguishable, the
lines are much broader. Furthermore, there is no detectable density
gradient or rotational pattern visible. In addition, the veil seen
between the two smaller pillars in M16 \citep{1996AJ....111.2349H} poses
a real challenge. Even the sophisticated simulations of
\citet{Mackey:2010uq} do not reproduce it. In our simulations (see
  e.g. Fig. \ref{fig_M7}) veils arise naturally due to the turbulent
  motions. This is a strong indication that the pillars in M16 or in
  NGC 7822 are produced by the interplay of pre-existing turbulent
  structures and ionizaing radiation.

\section{Conclusions and Discussion}
\label{Conclusions}
With iVINE, a tree-SPH code including ionization, we perform a
parameter study on the formation of pillar-like structures and
triggered star formation in HII regions. First we show that our simulations are
converged and the choice of boundary conditions does not affect the
outcome significantly (\S \ref{RES_res}). After that, we show that
ionizing radiation imposed on a turbulent molecular cloud can result
in the formation of pillar like structures which resemble observed
pillars in size, mass, density as well was velocity structure.  
 Especially the rotational pattern observed in
several pillars indicates that these where formed by the
ionization of a turbulent cloud and not by the RDI of preexisting clumps.

We thus conclude:
\begin{enumerate}
\item Varying the turbulent Mach number between $1.5-12.5$ changes the
  morphology. If the Mach number is too low, there are no pre-existing
  structures that can be enhanced, if it is too high, the structures
  disperse quickly. The most favorable regime for the formation of
  pillars is Mach $4-10$. However, the physical quantities such as the
  mass assembled and the column density are only weekly dependent on
  the Mach number (see \S \ref{RES_Mach}).
\item The formation of pillar-like structures in the case with an
  favorable Mach number critically depends on the
  initial density contrast. Structures and therefore stars only form
  if the density contrast is lower than the temperature contrast
  between the hot and the cold gas:
  $\Delta\rho_\mathrm{init}\le\frac{2T_\mathrm{ion}}{T_\mathrm{nion}}$
  (see \S \ref{RES_T}).
\item The evolution of the ionized mass, density and the front
  position in a turbulent medium under the influence of
  different initial fluxes is remarkably similar to the evolution in a
  homogeneous case. Thus, the size and density inside a HII-region
  solely depends on the initial flux, density and the time since the
  ignition of a star or the distance from the star and globally follows the
  simplified analytical prescription (see \S \ref{RES_flux}).
\item The density of the resulting pillars is determined by a pressure
  equilibrium between the hot and the cold gas. Therefore, the
  expected density of the structures can be calculated as well (see \S
  \ref{RES_general}).
\item The size of the evolving structures critically depends on the
  driving modes of the turbulence. Smaller driving modes lead to
  smaller structures. In our simulations the relation is roughly
  $d_\mathrm{pil}\approx x_\mathrm{driving}/40$ (see \S
  \ref{RES_scale}).
\item Core and star formation is likely to occur. The higher the mass
  in the structures and the higher the initial flux, the earlier cores
  form (see \S \ref{RES_SF}).
\end{enumerate}
Combining 3) and 4) allows us directly to determine the density of the
forming structures as function of the initial mean density of the medium,
the flux of the source, and the time since the ignition of the source
or the position of the ionization front (see Eq. \ref{rho_pred}).

One has to keep in mind that our approach is simplified. 
First of all, no scattering is taken into account. Once a
electron recombines, the emitted photon is assumed to be absorbed in
the direct neighborhood (on the spot approximation). Thus, the 
reheating of shadowed regions by the adjacent hot gas is not taken
into account. How much this affects the formation of
pillars is the topic of ongoing research (Ercolano \& Gritschneder, in
prep). In addition, we focus on atomic hydrogen only, which
makes it impossible to follow the precise temperature evolution as well as the
photodissociation regions (PDRs). On the other hand, our simulations
indicate that the pillars are in pressure equilibrium with the hot
gas. Therefore, the PDRs might be transition regions comparable to a
thin shock layer which is not resolved. Furthermore, we do not take
magnetic fields into account. These might have implications on the
global shapes of the HII region \citep[see
e.g.][]{2007ApJ...671..518K}. Nevertheless, we are able to 
reproduce the cork-screw morphologies in the pillars which were
sometimes attributed to magnetic fields (see \S
\ref{OBS_vel}). Another aspect we neglect 
are stellar winds. Although there is clear observational evidence
the winds of a massive star interact with the surrounding ISM
(e.g. \citealt{Westmoquette:2010fj}), it is up to now still unclear how effectively
they affect its surroundings. From our simulations we
would estimate that stellar winds are of minor importance, maybe
mainly enhancing the shock front as soon as a lower density in the hot gas
allows for the effective driving of a stellar wind.

Altogether, our simulations are able to reproduce even the detailed
fine-structure of the pillars that have been observed with high resolution.
In addition, we find that the observed line of sight profiles allow
for a clear distinction between the radiation driven implosion
scenario and the 'radiative round-up' presented here. Current
observations are in favor of our 'radiative round-up'
mechanism. Besides, our simulations give a deeper insight on the tight 
correlation between the parental molecular clouds size, density and
turbulence and the structures excavated by the ionizing radiation. The
ionization acts as a magnifying glass, revealing the condition of the
molecular cloud previous to the ignition of the massive star.

\begin{acknowledgements}
We thank the referee for valuable comments which helped to
improve the manuscript. 
This research was founded by the DFG cluster of excellence
'Origin and Structure of the Universe'. MG acknowledges additional
funding by the China National Postdoc Fund Grant No. 20100470108 and
the National Science Foundation of China Grant No. 11003001. 
 All simulations were performed on a SGI Altix 3700 Bx2 supercomputer
 that was partly funded by the DFG cluster of excellence 'Origin and Structure of the Universe'.
\end{acknowledgements}

\bibliographystyle{apj}
\bibliography{references}

\clearpage
\end{document}